\documentclass[12pt]{spieman}  
\usepackage{amsmath,amsfonts,amssymb}
\usepackage{graphicx}
\usepackage{setspace}
\usepackage{tocloft}

\usepackage{epstopdf}
\usepackage{lscape}
\usepackage{pdflscape}
\usepackage{pgfgantt}
\usepackage{rotating}
\usepackage{longtable}
\usepackage{multirow}

\title{Towards accurate quantitative photoacoustic imaging: learning vascular blood oxygen saturation in 3D}

\author[a,*]{Ciaran Bench}
\author[b,c]{Andreas Hauptmann}
\author[a]{Ben Cox}
\affil[a]{University College London, Department of Medical Physics and Biomedical Engineering, Gower Street, London, UK, WC1E 6BT}
\affil[b]{Research Unit of Mathematical Sciences, University of Oulu, Pentti Kaiteran katu 1, 90014, Oulu, Finland}
\affil[c]{University College London, Department of Computer Science, Gower Street, London, UK, WC1E 6BT}

\cftpagenumbersoff{figure}
\cftpagenumbersoff{table} 
\begin{document} 
\maketitle

\begin{abstract}
\textbf{Significance:} 2D fully convolutional neural networks have been shown capable of producing maps of sO$_2$ from 2D simulated images of simple tissue models. However, their potential to produce accurate estimates \textit{in vivo} is uncertain as they are limited by the 2D nature of the training data when the problem is inherently 3D, and they have not been tested with realistic images. 


\textbf{Aim:} To demonstrate the capability of deep neural networks to process whole 3D images and output 3D maps of vascular sO$_2$ from realistic tissue models/images.

\textbf{Approach:} Two separate fully convolutional neural networks were trained to produce 3D maps of vascular blood oxygen saturation and vessel positions from multiwavelength simulated images of tissue models.

\textbf{Results:} The mean of the absolute difference between the true mean vessel sO$_2$ and the network output for 40 examples was 4.4\% and the standard deviation was 4.5\%.

\textbf{Conclusions:} 3D fully convolutional networks were shown capable of producing accurate sO$_2$ maps using the full extent of spatial information contained within 3D images generated under conditions mimicking real imaging scenarios. This work demonstrates that networks can cope with some of the confounding effects present in real images such as limited-view artefacts, and have the potential to produce accurate estimates \textit{in vivo}.
\end{abstract}


{\noindent \footnotesize\textbf{*}Ciaran Bench,  \linkable{ciaran.bench.17@ucl.ac.uk} }

\begin{spacing}{1}   

\section{Introduction}
\label{sect:intro}  
Blood oxygen saturation (sO$_2$) is an important physiological indicator of tissue function and pathology. Often, the distribution of oxygen saturation values within a tissue is of clinical interest, and therefore, there is a demand for an imaging modality that can provide high resolution images of sO$_2$. For example, there is a known link between poor oxygenation in solid tumor cores and their resistance to chemotherapies, thus images of tumor blood oxygen saturation could be used to help stage cancers and monitor tumor therapies \cite{tomaszewski2017oxygen,ron2019volumetric}. Some imaging modalities have been shown capable of providing limited information about or related to sO$_2$ in tissue. Blood Oxygenation Level Dependent Magnetic Resonance Imaging (BOLD MRI), which is sensitive to changes in both blood volume and venous deoxyhaemoglobin concentration, can be used to image brain activity, but can not respond to changes in oxygen saturation \cite{ogawa1990brain}. Purely optical techniques, such as near-infrared spectroscopy (NIRS) and diffuse optical tomography (DOT), can be used to generate images of oxygen saturation \cite{villringer1993near,gibson2005recent}. However, because of high optical scattering in tissue, these modalities can only generate images with low spatial resolution beyond superficial depths.

Photoacoustic (PA) imaging is a hybrid modality that can be used to generate high resolution images of vessels and tissue at greater imaging depths than purely optical modalities \cite{beard2011biomedical}. PA image contrast depends on the optical absorption of the sample, so images of well-perfused tissues and vessels can in principle be used to generate images of sO$_2$ with high specificity. However, unlike strictly optical techniques, information about the contrast in PA images is carried by acoustic waves that can propagate from deep within a tissue to its surface undergoing little scattering. 

In the ideal case of a perfect acoustic reconstruction, the amplitude of a voxel in a PA image can be described by
\begin{equation}
p_0(\textbf{x},\lambda) = \mu_a(\textbf{x},\lambda)\Gamma(\textbf{x})\Phi(\textbf{x},\lambda; \mu_a, \mu_s, g),
\end{equation}
where $\textbf{x}$ is the voxel's location within the sample, $\lambda$ is the optical wavelength, $\mu_a$ is the optical absorption coefficient, $\mu_s$ is the scattering coefficient, g is the optical anisotropy factor, $\Gamma$ is the PA efficiency (assumed here to be wavelength independent), and $\Phi$ is the light fluence. Images of sO$_2$ may only be recovered if the sample's absorption coefficients (or at least the absorption coefficient scaled by some wavelength independent constant, such as $\mu_a(\textbf{x},\lambda)\Gamma(\textbf{x})$) can be extracted from each image. In the hypothetical case where the sample's fluence distribution is constant with wavelength, a set of PA images acquired at multiple wavelengths automatically satisfy this requirement. However, because the optical properties of common tissue constituents are wavelength dependent, this condition is never met in \textit{in vivo} imaging scenarios \cite{hochuli2019estimating}. In general, knowledge of the fluence distribution throughout the sample at each excitation wavelength is required to accurately image sO$_2$ \cite{cox2012quantitative}. If an accurate fluence estimate is available, then an image of the sample's relative optical absorption coefficient at a particular wavelength can be obtained by performing a voxel-wise division of the image by the corresponding fluence distribution, as described by
\begin{equation}
\frac{p_0(\textbf{x},\lambda)}{\Phi(\textbf{x},\lambda)} = \mu_a(\textbf{x},\lambda)\Gamma(\textbf{x}).
\end{equation}

In some cases it might be possible to measure an estimate of the fluence using an adjunct modality \cite{hussain2016quantitative}, but more commonly, attempts have been made to model the fluence. However, because the optical properties of a tissue sample are usually not known before imaging (the only reason the fluence is estimated at all is so that unknown information about the sample's optical absorption coefficient can be recovered from the image data), it is difficult to model the fluence distribution. A variety of techniques have been developed to recover tissue absorption coefficients from PA images without total prior knowledge of the tissue's optical properties. Progress towards solving this problem can be summarized into three key phases. In the first phase, 1D analytical fluence models were used to estimate the fluence by taking advantage of assumed prior knowledge of some of the sample's optical properties, or by extracting the optical properties of the most superficial layers from image data \cite{carome1964generation,cross1987time,cross1988ablative,guo2010calibration,Deng2016,Kim2011}. In the latter case, the effective attenuation coefficient of the most superficial tissue layer (assumed to be optically homogeneous) is usually estimated by fitting an exponential curve to the decay profile of the image amplitude above the region of interest (e.g. a blood vessel).

In the next phase, sample optical properties were recovered using iterative error minimisation approaches \cite{fonseca2017three,cox2006two,buchmann2020quantitative,buchmann2019three}. With these techniques, knowledge of the underlying physics is used to formulate a model of image generation. The set of model parameters (which might include the concentrations of deoxyhaemoglobin and oxyhaemoglobin in each voxel) that minimizes the error between the images generated by the model and the experimentally acquired images are treated as estimates of the same parameters in the real images. This technique is only effective when the model of image generation is able to generate a set of simulated images very similar to the real set of images when the correct values for the chromophore concentrations are estimated. This is only possible when the image generation model is able to accurately model image acquisition in the real system. In practice, accurate models of image generation are challenging to formulate as not all aspects of the data acquisition pathway are fully characterized. Therefore, this technique has not yet been shown to be a consistently accurate method for imaging sO$_2$ in tissue. Both iterative error-minimisation and analytical techniques may require significant \textit{a priori} knowledge of sample properties, such as all the different constituent chromophore types. This information is not always available when imaging tissues \textit{in vivo}, and thus this requirement further reduces their viability as techniques for estimating sO$_2$ in realistic imaging scenarios.
The recent emergence of a third phase has introduced data-driven approaches to solving the problem \cite{grohl2019estimation,yang2019eda,luke2019net,durairaj2020unsupervised,chen2020deep,yang2019quantitative, kirchner2018context, grohl2018confidence, cai2018end}. With these approaches, generic models are trained to output images of sO$_2$ or optical properties by processing a set of examples \cite{arridge2019solving}. These data-driven models find solutions without significant \textit{a priori} knowledge of sample properties, and do not require the formulation of an image generation model using assumed prior knowledge of all the aspects related to image acquisition. Techniques based on data-driven models, such as Deep Learning, have been used to estimate sO$_2$ from 2D PA images of simulated phantoms and tissue models \cite{grohl2019estimation,yang2019eda,luke2019net,chen2020deep,yang2019quantitative,cai2018end}. Fully-connected feed-forward neural networks have been trained to estimate the sO$_2$ in individual image pixels given their PA amplitude at multiple wavelengths \cite{grohl2019estimation}. Because the fluence depends on the 3D distribution of absorbers and scatters, a pixel-wise approach does not use the full information available in an image. Encoder-decoder type networks, capable of utilising spatial as well as spectral information, have been trained to process whole multiwavelength 2D images of 2D tissue models \cite{luke2019net,chen2020deep,yang2019quantitative,cai2018end}, or 2D images sliced from more realistic 3D tissue models featuring reconstruction artefacts \cite{yang2019eda}, and output a corresponding 2D image of the sO$_2$/optical absorption coefficient distribution. Although 2D convolutional neural networks can take advantage of spatial information to improve estimates of sO$_2$, networks trained on 2D images sliced from 3D images are missing information contained in other image slices that might improve their ability to learn a fluence correction. Therefore, it is important to show that networks can take advantage of all four dimensions of information from a multiwavelength PA image dataset to estimate sO$_2$. In addition to supervised learning, an unsupervised learning approach has been used to identify regions containing specific chromophores (such as oxyhaemoglobin and deoxyhaemoglobin) in 2D simulated images \cite{durairaj2020unsupervised}. The technique has not yet been used to estimate sO$_2$, and has only been tested on a single simulated phantom lacking a complex distribution of absorbers and scatterers that would normally be found in \textit{in vivo} imaging scenarios. 


As we aim towards developing a technique for estimating 3D sO$_2$ distributions from \textit{in vivo} image data, a more robust demonstration of a data-driven technique’s ability to acquire accurate sO$_2$ estimates by processing whole 3D images of realistic tissue models is desired. 
We trained two encoder-decoder type networks with skip connections to 1) output a 3D image of vascular sO$_2$, and 2) output an image of vessel locations from multiwavelength (784 nm, 796 nm, 808 nm, 820 nm) images of realistic vascular architectures immersed in three-layer skin models, featuring noise, and reconstruction artefacts. Details about how the simulated images were generated are described in Section \ref{sec:Image_gen}. Section \ref{sec:Traininng_param} describes the network architecture and details about the training process. Section \ref{sec:Results} describes the results.

\section{Generating Simulated Images}
\label{sec:Image_gen}
Ideally, a network trained to estimate sO$_2$ from \textit{in vivo} images would be capable of generating accurate estimates from a wide range of tissue samples with varying optical properties and distributions of vessels. Additionally, the network should be able to do this despite the presence of reconstruction artefacts and noise. This section describes each step involved in the generation of the simulated images used in this study.

\subsection{Tissue Models}
\begin{figure}[tbp]
\centering
\includegraphics[width=.6\columnwidth]{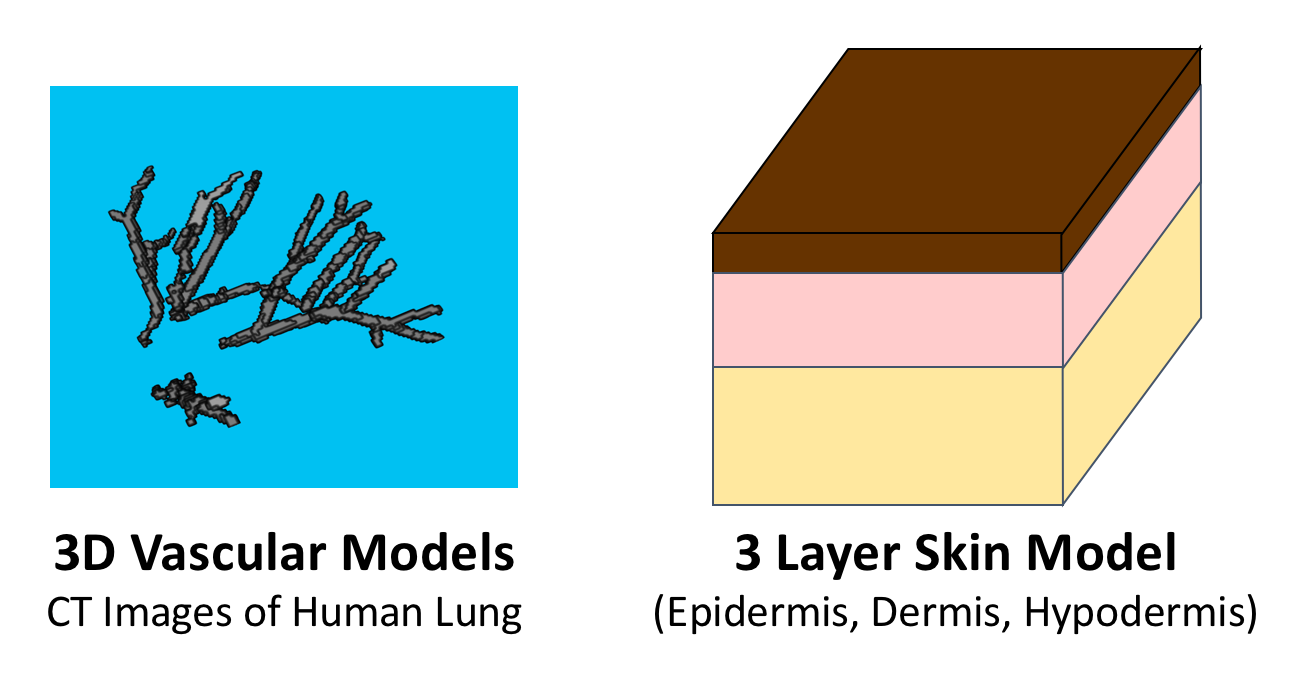}
\caption{Left: Example of a 3D vessel model (acquired from CT images of human lungs) used to construct 3D tissue models. Right: Schematic of three-layer skin model used to construct tissue models.}
\label{fig:Vessel_models}
\end{figure}

A set of several hundred tissue models, each featuring a unique vascular architecture and distribution of optical properties, were generated by immersing 3D vessel models acquired from CT images of human lung vessels into 3D, three-layer skin models (some examples are shown in Fig. \ref{fig:Vessel_models}) \cite{via_group,hauptmann2018model}. Each skin model consisted of an epidermis, dermis and hypodermis layer. The thickness of each skin layer (epidermis: 0.1 - 0.3 mm , dermis: 1.3 mm - 2.9 mm, hypodermis: 0.8 mm - 2.6 mm), and the optical absorption properties of the epidermis and dermis layers were varied for each tissue model. A unique tissue model was generated for each vascular model. The equations used to calculate the optical properties of each skin layer and the vessels at each excitation wavelength (784 nm, 796 nm, 808 nm, 820 nm) are presented in Table \ref{table:Skinproperties} in Appendix \ref{sec:Optical_properties}. These wavelengths were chosen as they fell within the NIR and data was available for all skin layers at these wavelengths. The absorption properties of the epidermis layer of each tissue model were determined by choosing a random value for the melanosome volume fraction that was within expected the physiological range. The absorption properties of the dermis layer were determined by choosing random values for the blood volume fraction and dermis blood sO$_2$ within the expected physiological range. For each tissue model, each independent vascular body was randomly assigned one of three randomly generated sO$_2$ values between $0\%$ and $100\%$. The PA efficiency throughout the tissue was set to one with no loss of generality. 

\subsection{Fluence Simulations}
The fluence in each tissue model at each excitation wavelength was simulated with MCXLAB, a MATLAB package that implements a Monte Carlo (MC) model of light transport (considered the gold standard for estimating the fluence distribution in tissue models) \cite{fang2009monte}. Fluence simulations were run with 10$^{9}$ photons, the maximum number of photons that could be used to generate 1024 sets of images in approximately one week using a single NVIDIA Titan X Maxwell GPU with 3072 CUDA cores and 12 GB of memory. A large number of photons were used in order to reduce the MC variance to the point where it no longer contributed significantly to the noise in the simulated data. Noise was subsequently added in a systematic way to the simulated time series, as described in Section \ref{sec:Acoustic}.

MC simulations were run with voxel sidelengths of 0.1 mm, and simulation volumes with dimensions of 40 x 120 x 120 voxels. The fluence was calculated from the flux output from MCXLAB by integrating over time using timesteps of 0.01 ns for a total of 1 ns, which was sufficient to capture the contributions from the vast majority of the scattered photons. A truncated Gaussian beam with a waist radius of 140 voxels, with its centre placed on the centre of the top layer of epidermis tissue was used as the excitation source for this simulation. Photons exiting the domain were terminated. The fluence simulations were not scaled by any real unit of energy, as images were normalized before inserting them into the network. Each fluence distribution was multiplied pixel-wise by an image of the tissue model's corresponding optical absorption coefficients to produce images of the initial pressure distribution at each excitation wavelength.

\subsection{Acoustic Propagation and Image Reconstruction}
\label{sec:Acoustic}
Simulations of the acoustic propagation of the initial pressure distributions from each tissue model, the detection of the corresponding acoustic pressure time series at the tissue surface by a detector with a planar geometry, and the time reversal reconstruction of the initial pressure distributions from these times series were executed in k-Wave \cite{treeby2010k}. Simulations were designed with a grid spacing of 0.1 mm, dimensions of 40 x 120 x 120 voxels, and a perfectly matched layer of 10 voxels surrounding the simulation environment. Each tissue model was assigned a homogeneous sound speed of 1500 m/s. 2D planar sensor arrays are often used to image tissue \textit{in vivo}, as it is a convenient geometry for accessing various regions on the body \cite{plumb2018rapid,huynh2016photoacoustic,huynh2017sub}. A sensor array with a planar geometry was used in this study to mimic conditions expected in real imaging scenarios. A 2D planar sensor mask covering the top plane of the tissue model was used to acquire the time series data. Because of its limited-view geometry, the sensor array will detect less pressure data emitted from deeper within the tissue, as these regions will subtend a smaller angle with the sensor. As a result, the reconstruction will have limited-view artefacts which will become more pronounced with depth \cite{xu2004reconstructions}.

To avoid the large grid dimensions that would be required to capture the abrupt change in the acoustic pressure distribution at the tissue surface, and consequently the long simulation times, the background signal in the top three voxel planes was set to zero. This has a similar effect to the bandlimiting of the signal during measurement that would occur in practice, and has no effect on the simulation of the artefacts around the vessels due to the limited detection aperture. Furthermore, in experimental images the superficial layer is often stripped away to aid the visualisation of the underlying structures. Similar approaches have been used to improve sO$_2$ estimates generated by 2D networks. In \cite{luke2019net}, the ten most superficial pixel rows were removed from images before training to ensure that features deeper within the tissue (and therefore dimmer than the comparatively bright superficial layers) were more detectable. Similarly, superficial voxel layers were removed from images in \cite{yang2019eda} to improve the accuracy of sO$_2$ estimates.


Noise was added to each datapoint in the simulated pressure time series by adding a random number sampled from a Gaussian distribution with a standard deviation of 1\% of the maximum value over all time series data generated from the same image, resulting in realistic SNRs of about 21 dB. Details of how this noise test was carried out, and how the SNR was calculated are provided in Appendix \ref{sec:Noise_test}. 

\section{Network Architecture and Training Parameters}
\label{sec:Traininng_param}
\begin{figure*}

  \includegraphics[width=\textwidth]{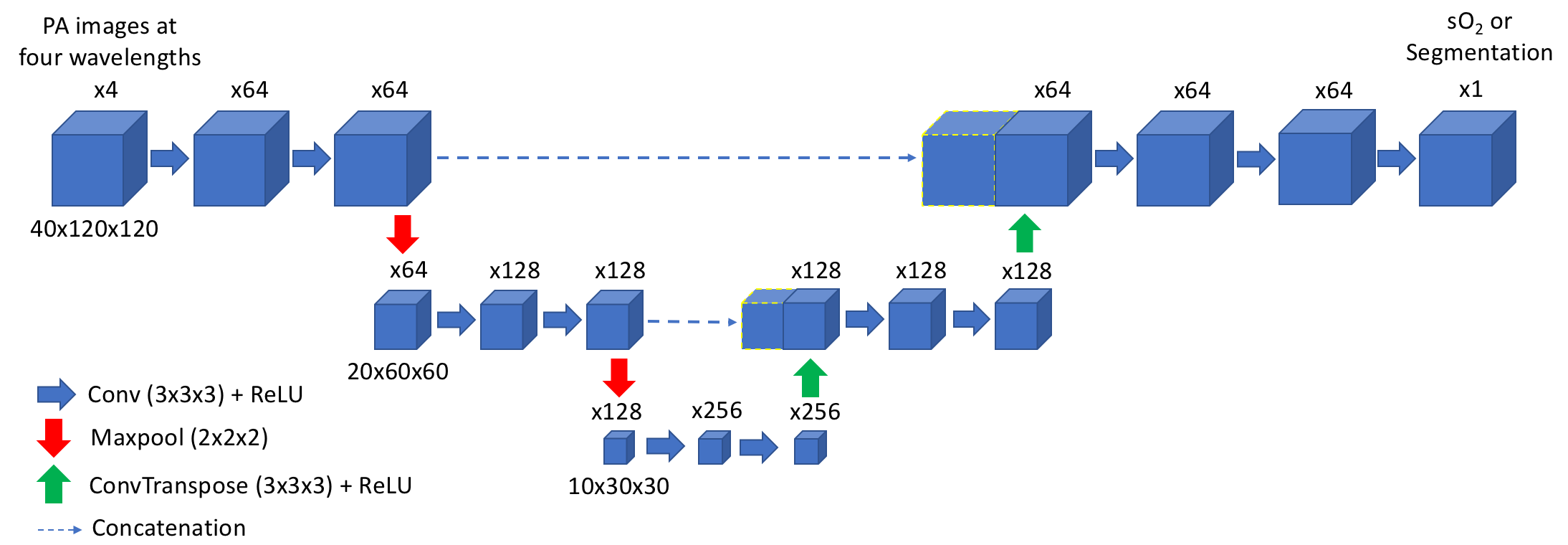}
  \caption{EDS network architecture. Blocks represent feature maps, where the number of feature maps generated by a convolutional layer is written above each block. Red arrows denote convolutional layers, green arrows denote maxpooling layers, dashed lines denote skip connections. }
  \label{fig:architecture}
\end{figure*}

A convolutional encoder-decoder type network with skip connections (EDS) (shown in Fig. \ref{fig:architecture}, and denoted as network $A$) was trained to output an image of the sO$_2$ distribution in each tissue model from 3D image data acquired at four wavelengths. Another network, network $B$, was assigned an identical architecture to network $A$ and was trained to output an image of vessel locations from the image sets (thereby segmenting the vessels). An EDS architecture was chosen for each task, as they have been shown to perform well at image-to-image regression tasks (i.e. tasks where the input data is a set of images, and the output is an image) \cite{ronneberger2015u}. The architecture takes reconstructed 3D images of a tissue model acquired at each excitation wavelength as an input. The multi-scale nature of the network allows it to capture information about features at various resolutions, and use image context at multiple scales \cite{piao2019accuracy,zeiler2013stochastic,jaderberg2015spatial}. The network's skip connections improve the stability of training, and help retain information at finer resolutions. Finally, the network outputs a single 3D feature map of the sO$_2$ distribution, or the vessel segmentation map.



An EDS network was trained to segment vessel locations because, although prior knowledge of the locations of vessels in the images was available for this \textit{in silico} study, this information will not always be available when imaging tissues \textit{in vivo}. Some technique for segmenting vessel positions from images is needed to enable the estimation of mean vessel sO$_2$ values from the output map. Therefore, a vessel segmentation network was trained to show that neural networks can be used to acquire accurate mean vessel sO$_2$ estimates without prior knowledge of vessels positions. As will be discussed in Section \ref{sec:Results}, the output of the segmentation network also provides some information about where estimates in the output sO$_2$ map may be more uncertain. This information can be used to improve mean vessel sO$_2$ estimates by disregarding values from these regions. Two separate networks were trained for each task to limit additional bias in the learned features that would arise from training a single network to learn both tasks simultaneously. In \cite{luke2019net}, two different loss functions were used in a single network trained to produce both an image of the vascular sO$_2$ distribution and an image of vessel locations. A different loss function was used for each task/branch of the network, where each function was arbitrarily assigned equal weights. Training two separate networks has the benefit that it removes the need to assign arbitrary weights to multiple loss functions that may be used to train a single network.

\subsection{Training Parameters}
\begin{figure}
\centering
  \includegraphics[width=.5\columnwidth]{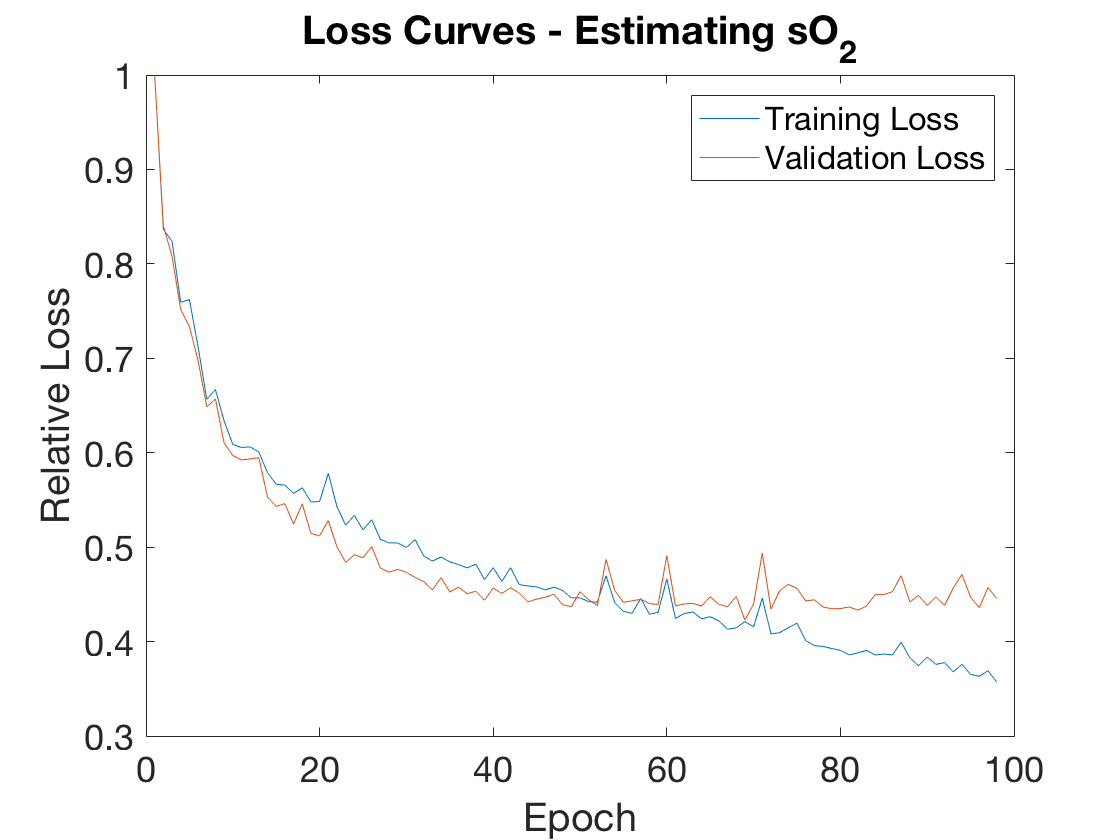}
  
  \caption{Relative loss curves ($\epsilon_A$) for the sO$_2$ estimating network.}
  \label{fig:Loss_curves}
\end{figure}

Networks $A$ and $B$ were trained with 500 sets of images, corresponding to 500 different tissue models. An image of the true sO$_2$ distribution of the vessels was used as the ground truth for network $A$. A binary image of true vessel locations was used as the ground truth for network $B$. Network $A$ was trained for 98 epochs (loss curve shown in Fig. \ref{fig:Loss_curves}), while network $B$ was trained for 84. Both networks were trained with a batch size of 5 image sets, a learning rate of $10^{-4}$, and with Adam as the optimizer. Training was terminated with an early stopping approach, using a validation set of 5 examples. The networks were trained with the following error functionals $\epsilon(\theta_A)$, and $\epsilon(\theta_B)$ (the norm of the squared difference between the network outputs and the ground truth images):
\begin{equation}\label{error_function}
    \epsilon_{\theta_A} = ||A_{\theta_A}(p_{0}(x,\lambda)) - sO_{2}^{true}(x) ||^{2} _{2},
\end{equation}
and,
\begin{equation}\label{error_function}
    \epsilon_{\theta_B} = ||B_{\theta_B}(p_{0}(x,\lambda)) - seg^{true}(x) ||^{2} _{2},
\end{equation}
where $p_{0}(x,\lambda)$ are the multiwavelength images of each tissue model, $sO_{2}^{true}(x)$ and $seg^{true}(x)$ are the ground truth sO$_2$ and vessel segmentation images, and $\theta_A$ and $\theta_B$ are the network parameters. Once trained, the networks $A_{\theta_A}$, and $B_{\theta_B}$ were evaluated on 40 test examples.

\subsection{Output Processing}
\label{sec:Image_processing}

The mean sO$_2$ of each vascular body was calculated using the voxels in each body that the segmentation network was confident contained vessels.

First, the indices associated with each major body in the segmentation network output ($V(x)$ where $x$ denotes the voxel index) were identified with the following method.

The output of the segmentation network $V(x)$ (where $x$ denotes the voxel index) was thresholded so all voxels with intensities $<0.2$ were set to zero, producing a new image $V'(x)$. This was done to remove small values which connected all the vessels into one large body. Then, the indices associated with each major body in $V'(x)$ were identified using the \verb bwlabeln() MATLAB function, generating a labelled image $L(x)$, where all the voxels belonging to each independent connected body were assigned the same integer value, and each body in the image was assigned a unique value to be identified by (e.g. all the voxels belonging to the largest body were assigned a value of one, all the voxels in the second largest body were assigned a value of two and so on). 

Then, $V'(x)$ was thresholded so all voxels with intensities $<1$ were set to zero, producing a new image $V''(x)$. This was done to isolate voxels where the network was confident that vessels were present. 

All the voxels in $L(x)$ that now had values of zero in $V''(x)$ were also set to zero, producing a new image $L_v(x)$. Bodies that were once a part of the same body before this thresholding step may now be in separate bodies, however, their voxel ID retains information about which body they originally belonged to. This allows for the mean sO$_2$ in each major vessel body to be calculated despite the thresholding (which removed voxels with low values in the segmentation network output) breaking up voxels which were once apart of the same body.

The mean sO$_2$ of the voxels sharing the same integer value in $L_v(x)$ were calculated using the corresponding values in the output of the sO$_2$ estimating network. The ground truth mean sO$_2$ of the voxels sharing the same integer value in $L_v(x)$ were calculated using the values from the ground truth sO$_2$ distribution.



\section{Results and Discussion}
\label{sec:Results}

\begin{figure}
  
  \phantom{a}{\Large a.\par}
  
  \includegraphics[width=.5\columnwidth]{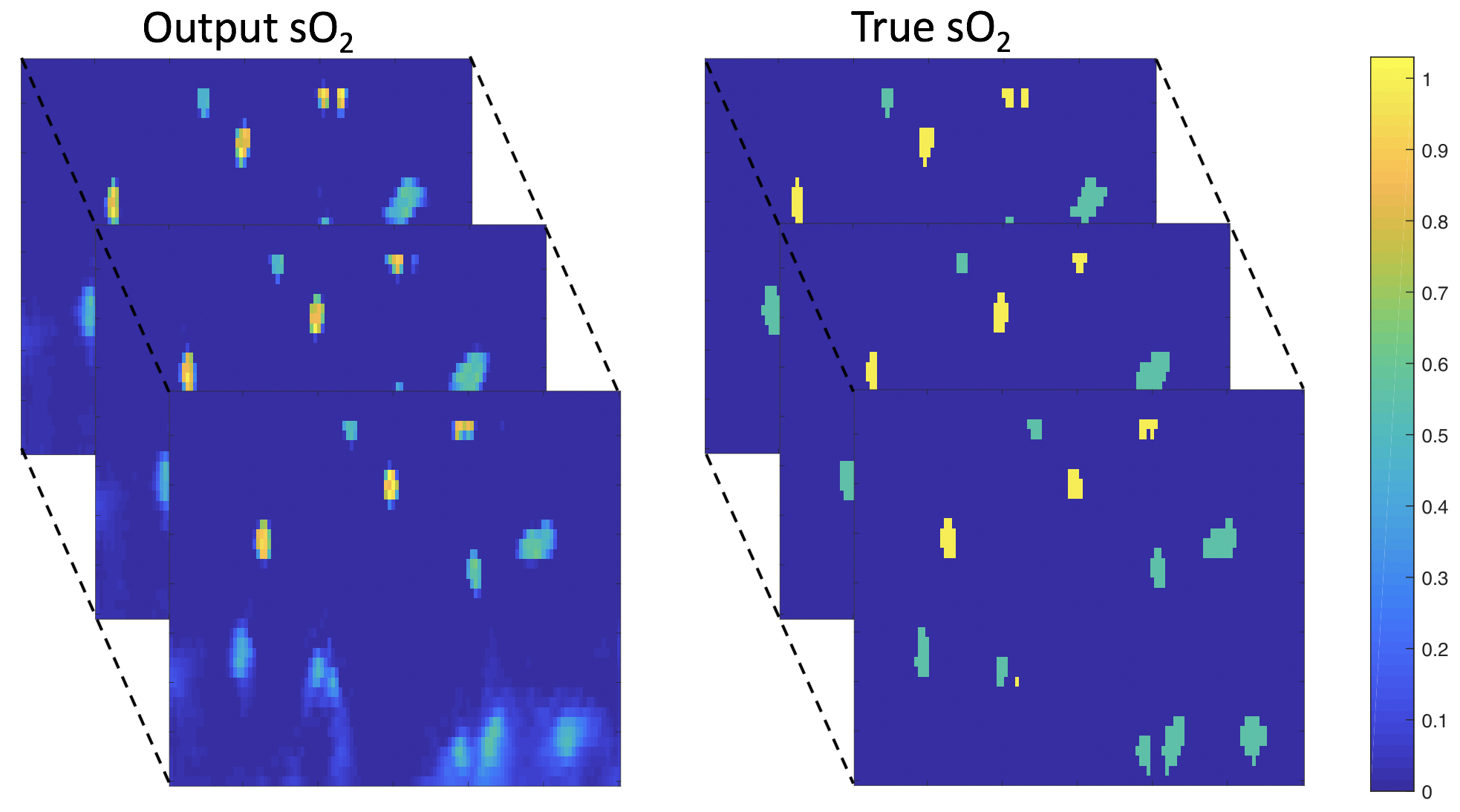}
  \includegraphics[width=.5\columnwidth]{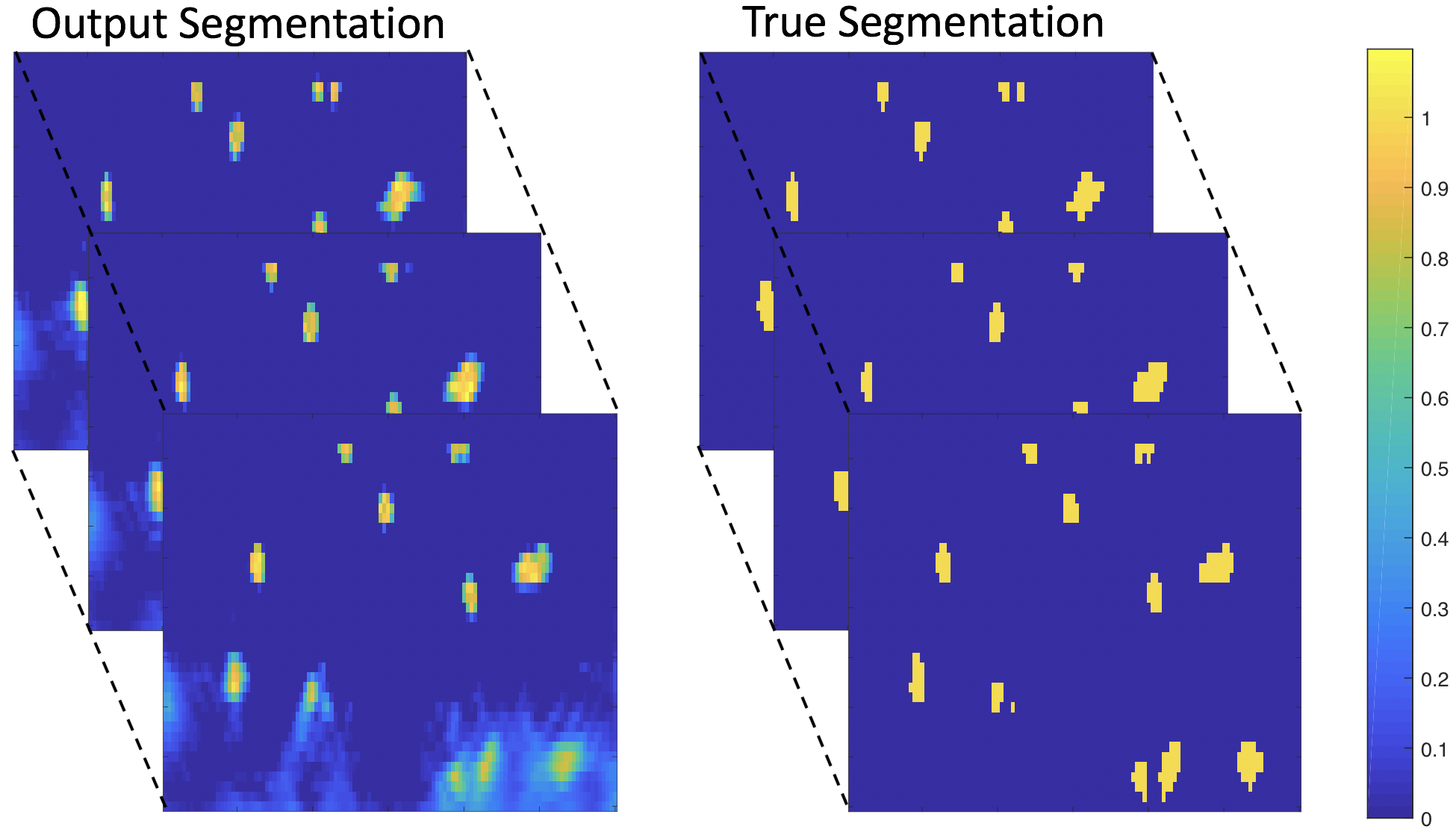}
  \phantom{a}{\Large b.\par}
  
  \includegraphics[width=.5\columnwidth]{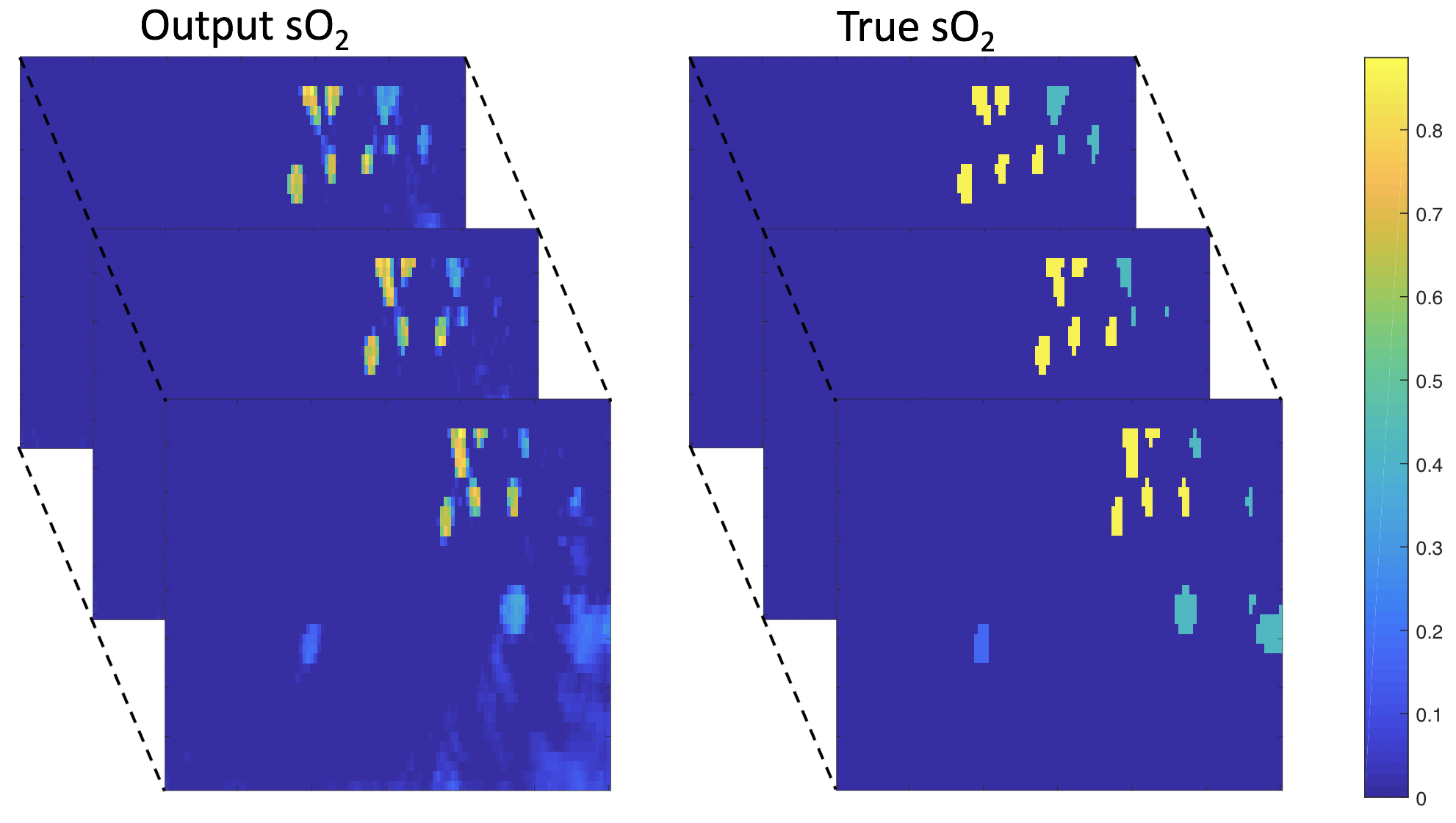}
  \includegraphics[width=.5\columnwidth]{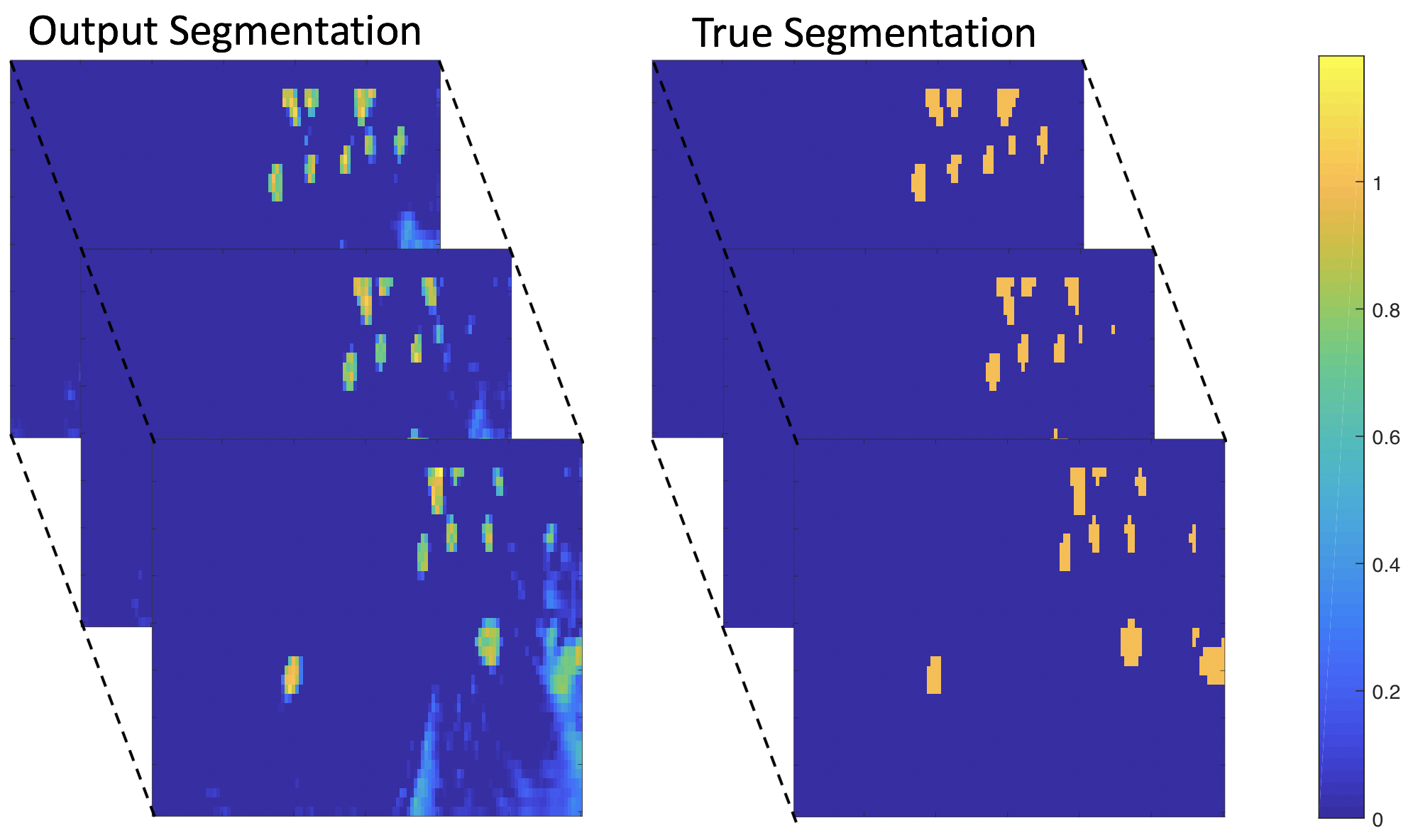}
  \caption{2D slices of 3D network outputs and corresponding ground truth sO$_2$ and vessel segmentation images for two different tissue models (labelled a and b). }
  \label{fig:Graph_result}
\end{figure}
\begin{figure}

  \begin{tabular}{ll}
  {\Large a.\par} & {\Large b.\par} \\
  \includegraphics[width=.4\columnwidth]{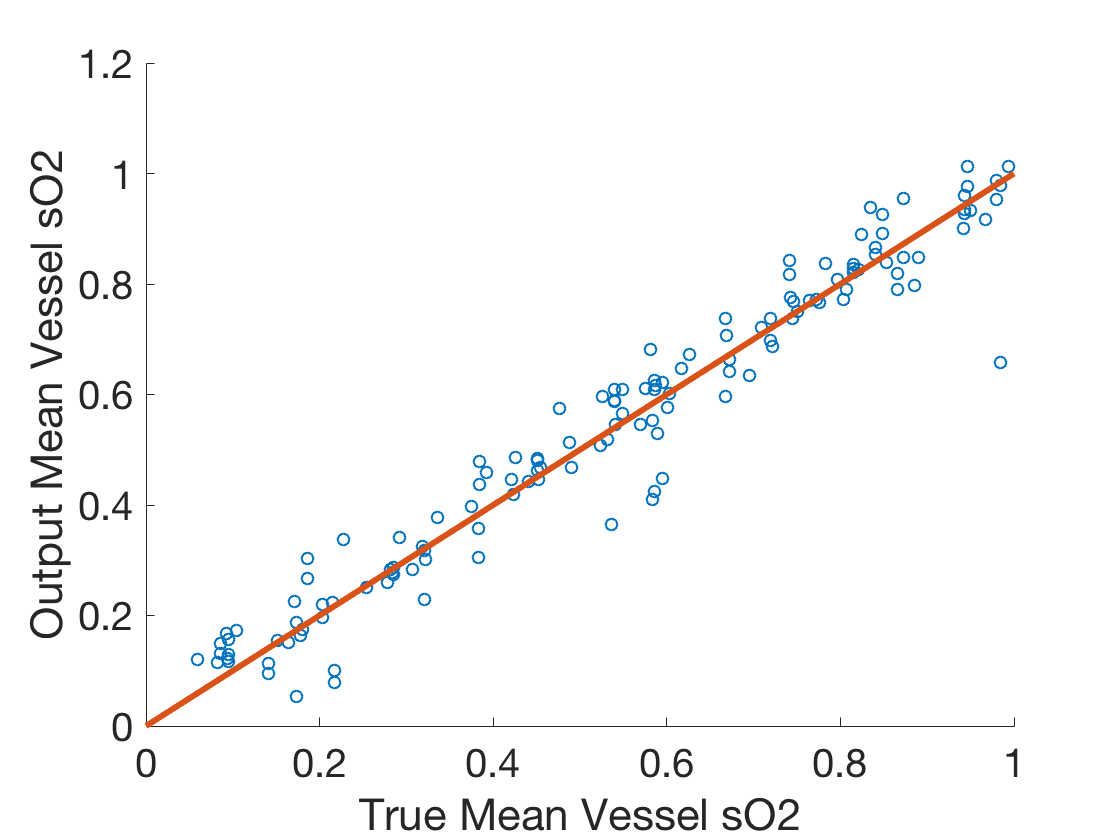}
  &
  
  \includegraphics[width=.4\columnwidth]{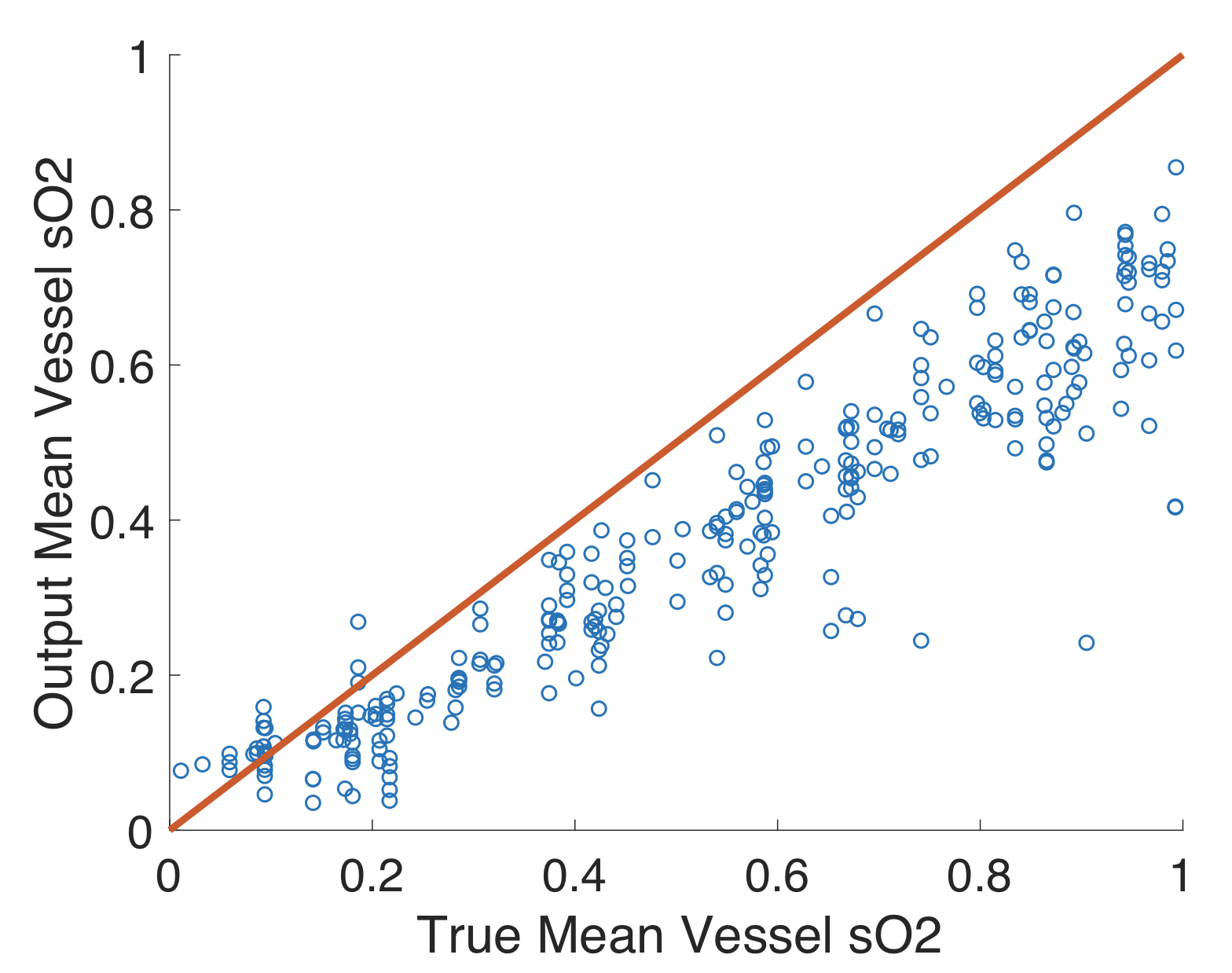}
  \end{tabular}
  \caption{a.) Plot of the output mean vessel sO$_2$ vs the true values for all the vessels in 40 tissue models not used for training, calculated with the voxels belonging to each vessel as determined by the segmentation network output. b.) Plot of the mean sO$_2$ values for the same 40 tissue models calculated using the voxels known to belong to each vessel as determined by the ground truth vessel positions. These plots show that using the output of the segmentation network in combination with the output of the sO$_2$-estimating network significantly improves the accuracy of the estimates.}
  \label{fig:plot_result}

\end{figure}

The 3D image outputs of both the sO$_2$-estimating and segmentation networks were processed in order to calculate the mean sO$_2$ in each major vessel body using only the sO$_2$ estimates from voxels that the segmentation network was confident contained vessels (the reason for using the segmentation output was so that the mean vessel sO$_2$ could be calculated without \textit{a priori} knowledge of vessel locations that might not be available in an \textit{in vivo} scenario). More details about how this process was performed are provided in Section \ref{sec:Image_processing}. The mean of the absolute difference between the true mean vessel sO$_2$ and the output mean vessel sO$_2$ over all 40 sets of images was 4.4\%, and the standard deviation of the absolute difference between the true mean vessel sO$_2$ and the output mean vessel sO$_2$ was 4.5\% (some 2D image slices taken from the networks' 3D outputs are shown in Fig. \ref{fig:Graph_result}, and a plot of all the estimates is provided in Fig. \ref{fig:plot_result}.a). Therefore on average, the predicted mean vessel sO$_2$ was within 5\% of the true value. The mean difference between the true mean vessel sO$_2$ and the output mean vessel sO$_2$ was -0.3\% with an standard deviation of 6.3\%. The typical error for a mean vessel sO$_2$ estimate was thus between -6.6\% to 6.0\%.

To assess the effect that using the output of the segmentation network may have had on the accuracy of the sO$_2$ estimates, the mean sO$_2$ of each vascular body in the network output was estimated using the voxels known to belong to each vessel, as opposed to the voxels assigned to each body by the segmentation network output. Curiously, the accuracy of the estimates decreased when the ground truth vessel voxels were used for calculating mean sO$_2$ values. Fig. \ref{fig:plot_result}.b shows a plot of the results over 40 tissue models. The mean of the absolute value of the offset between the true value and the network output was 16.6\%, the mean offset was 16.2\%, and the standard deviation of the offset was 11.5\%. This suggests that regions where the segmentation network confidently classified as belonging to a vessel corresponded to regions where the sO$_2$ network was more accurate. Even though both networks are trained separately, they share the same input data, network architecture, and are trained with the same loss function where only the distribution of values in the corresponding ground-truth varies (continuous versus binary). As such it is not surprising that the learned mapping properties are similar and complement each other. The L2 loss was used for training network B (as opposed to a binary classification loss function that would normally be used for a segmentation task) to ensure that the network outputs would retain more information about the uncertainty of estimates.



The accuracy of the output of both networks $A$ and $B$ decreased with the depth of the vessels, i.e. the distance from the detector array, as can be seen in Fig. \ref{fig:Graph_result}. There are a couple of reasons for why this might be the case. Firstly, image SNR decreases with depth as the amplitude of the signal decreases as fewer photons will propagate to/through the deeper tissue regions. Secondly, because the simulated sensor array has a limited-view planar geometry, it will detect less pressure data emitted from deeper within the tissue. The artefacts at greater depths are spread over more of space, introducing greater uncertainty as to the shape and location of the vessel. The output of both networks are least accurate in the deepest corners of each images, where the artefacts are the most significant.

Filters in a convolutional layer are the same wherever they are applied in the image - they are spatially invariant - and are therefore most suited to detecting features that are also spatially invariant. However, the limited-view artefacts are not; they are small close to the centre of the sensor array and become more significant further away. The multi-scale nature increases the receptive field and hence locality can be learned by the network. Nevertheless, we decided to limit the receptive field by using a slightly smaller network architecture than the classic U-Net, this way we retain uncertainty in the deeper tissue layers instead of introducing a learned bias.


The 3D results shown in Figs. \ref{fig:Graph_result} and \ref{fig:plot_result}.a are of comparable accuracy to results from other groups obtained by training 2D convolutional neural networks to process 2D images (lacking the presence of reconstruction artefacts) of simpler 2D tissue models \cite{luke2019net,chen2020deep,yang2019quantitative}. The technique presented here was not only able to handle more complex tissue models (the tissue models presented here feature more realistic vascular architectures, and multiple skin layers with varying thicknesses and optical properties), but also took as the input data 3D images featuring noise and reconstruction artefacts. Unlike networks trained on 2D images sliced from 3D images of tissue models (such as those used in \cite{yang2019eda}), the 3D networks were able to use information from entire 3D image volumes to generate estimates. Because the fluence distribution and limited-view artefacts are 3D in nature, learning 3D features is more efficient than trying to learn to represent 2D sections/slices through 3D objects with 2D feature maps. This increases their ability to produce accurate estimates in more complex tissue models.

Despite being more sophisticated than other tissue models used to date, the tissue models used here were nevertheless created with some simplifying assumptions. Each skin layer was assigned a planar geometry, where the value of the optical properties associated with each layer at each wavelength remained constant within each layer (e.g. the scattering coefficient of the epidermis was constant within the epidermis layer). Although the absorption coefficient of each layer was varied for each tissue model, the scattering coefficient of each skin type remained constant (but did vary with wavelength). Other experimental factors that can affect image amplitude, such as the directivity of the acoustic sensors, were not incorporated into the simulation pipeline. It remains to be seen the extent to which these assumptions will hold true when this network is applied to \textit{in vivo} data. To ensure that networks initially trained on simplified simulated images can output accurate estimates when provided real images, networks may have to be modified with transfer training, taking advantage of datasets of real images \cite{pan2009survey,hauptmann2018model,wirkert2017physiological}. Looking beyond the complexity of the tissue models, there are other more fundamental challenges that will make the application to living tissue non-trivial. In order to train a network using a supervised learning approach with \textit{in vivo} data (or even to validate any technique for estimating sO$_2$ \textit{in vivo}), the corresponding ground truth sO$_2$ distribution must be available. It is unclear as to how this information might be acquired, and this poses a significant challenge that must be overcome to realize or validate the application of the technique. As an intermediate step towards generating \textit{in vivo} datasets, blood flow phantoms with tuneable sO$_2$ could be used to generate labelled data in conditions mimicking realistic imaging scenarios \cite{vogt2019photoacoustic,gehrung2019development}. 
Although it is important to show that a network can cope with all the confounding effects present in real images of tissue, it is still interesting and important to know that the technique can cope with at least some of the challenges faced in such scenarios. This work provides an essential demonstration of the technique's ability to generate accurate 3D estimates from 3D image data despite the presence of some confounding experimental effects which distort image amplitude, and despite some variation in the distribution of tissue types and the distribution of vessels for each tissue model.

\section{Conclusions}
Data-driven approaches have been shown capable of recovering sample optical properties and maps of sO$_2$ from 2D PA images of fairly simple tissue models. However, because the fluence distribution  and limited-view artefacts are 3D, 2D networks are at a disadvantage as they must learn to represent 2D sections/slices through 3D objects with 2D feature maps. Networks that can process whole 3D images with 3D filters are more efficient as they can detect 3D features, and this increases their ability to produce accurate estimates in more complex tissue models. There may be cases where accurate sO$_2$ maps may only be generated with 3D network architectures. Therefore, to assess whether data-driven techniques have the potential to provide accurate estimates in realistic imaging scenarios, it is essential to demonstrate a neural network's ability to process 3D image data to generate sO$_2$ estimates. The capability of an EDS to generate accurate maps of vessel sO$_2$ and vessel locations from multiwavelength simulated images (containing noise and limited view artefacts) of tissue models featuring optically heterogeneous backgrounds (with varying absorption properties), and realistic vessel architectures was demonstrated. Regions where the segmentation output was confident in its predictions of vessel locations corresponded to more accurate regions in the sO$_2$-estimating network output. As a consequence, the accuracy of the network's mean vessel sO$_2$ estimates improved when the output of the segmentation network was used to determine vessel locations as opposed to the ground truth. In contrast to both analytical and iterative error-minimisation techniques, the networks were able to generate these estimates without total knowledge of each tissues' constituent chromophores, or an accurate image generation model - both of which would not normally be available in a typical \textit{in vivo} imaging scenario. This work shows that fully convolutional neural networks can process whole 3D images of tissues to generate accurate 3D images of vascular sO$_2$ distributions, and that accurate estimates can be generated despite some degree of variation in the distribution of tissue types, vessels, and the presence of noise and reconstruction artefacts in the data. 

\appendix

\section{Noise Test}
\label{sec:Noise_test}
Noise was incorporated into the simulated images by adding it the the simulated pressure time series before the reconstruction step. Noise was added to each datapoint in the simulated pressure time series by adding a random number sampled from a Gaussian distribution with a standard deviation of 1\% of the maximum value over all time series data generated from the same image, resulting in realistic SNRs of 20.9 dB, 21.3 dB, 21.4 dB and 21.4 dB for a set of images of a single tissue model simulated at 784 nm, 796 nm, 808 nm and 820 nm, respectively. The details of this measurement are described in the following section.
\begin{enumerate}
    \item A single tissue model was defined. 
    \item A fluence simulation was run 20 times (each run indexed with $r$) for each excitation wavelength ($\lambda$) with $10^{9}$ photons to produce $\Phi_r(x,\lambda)$, where $x$ indexes the voxels in the simulation output. The optical properties of the tissue model at each excitation wavelength were identical for all 20 runs. 
    \item A set of initial pressure distributions, $I_r(x,\lambda)$, were generated from the fluence simulations:
    \begin{equation}\label{eq:SNR_image_recon}
        I_r(x,\lambda) = \Phi_r(x,\lambda)\mu_a(x,\lambda)\Gamma.
    \end{equation}
    \item The emission and detection of pressure time series were simulated in k-Wave to generate simulated pressure time series $p_{r,j}(t,\lambda)$, where $j$ indexes each time series produced by the simulation, and $t$ is the simulation time (simulation parameters were identical to those outlined in Section \ref{sec:Acoustic}). 
    \item Some amount of noise $n_{r,j}(t,\lambda)$ was added to each point in each pressure time series:
    \begin{equation}\label{eq:SNR_noise}
        \hat{p}_{r,j}(t,\lambda) = n_{r,j}(t,\lambda)+p_{r,j}(t,\lambda),
    \end{equation}
    where $n_{r,j}(t,\lambda)$ was determined by sampling a random value from a Gaussian distribution with a standard deviation of $cn_{max}(\lambda)$, where $n_{max}(\lambda)$ is the max value of $p_{r,j}(t,\lambda)$ over all $j$ and $t$, for a given $\lambda$, and $r$ (i.e. the max value of all the time series for a given run at a given wavelength), while $c$ is the proportion of this value used to define the standard deviation. 
    \item The images were reconstructed in k-Wave with time-reversal, to produce $I_r^{Recon}(x,\lambda)$.
    \item The mean and standard deviation for each voxel for each wavelength over all 20 runs was calculated with: 
    \begin{equation}\label{eq:SNR_mu_2}
        \mu(x,\lambda) = \frac{1}{20}\sum^{20}_{r=1}I_r^{Recon}(x,\lambda),
    \end{equation}
    and
    \begin{equation}\label{eq:SNR_sigma_2}
        \sigma(x,\lambda) = \left(\frac{1}{19}\sum^{20}_{r=1}(I_r^{Recon}(x,\lambda)-\mu(x,\lambda))\right)^{\frac{1}{2}}.
    \end{equation}
     \item The SNR of each voxel at each wavelength was calculated with:
    \begin{equation}\label{eq:pixel_SNR}
        SNR(x,\lambda) = 20\log_{10}(\frac{\mu(x,\lambda)}{\sigma(x,\lambda)}).
    \end{equation}
    \item For each wavelength, the mean of the SNR values over all voxels $V$ was calculated with:
    \begin{equation}\label{eq:SNR_mu_t}
        \mu_{SNR}(\lambda) = \sum^{V}_{x=1}SNR(x,\lambda).
    \end{equation}

\end{enumerate}

Because the SNR depends on the optical properties of objects in the sample domain, the SNR will vary depending on the tissue model used for the test. Here, we only use a single tissue model with a single set of tissue properties to obtain some approximate idea of how much noise features in the simulated images.

\section{Optical Properties of Skin Layers}
\label{sec:Optical_properties}
The refractive index, anisotropy factor, optical absorption coefficient, and optical scattering coefficient of each tissue or chromophore are required to construct a tissue model for a MCXLAB simulation. Here, we tabulate expressions for computing the relevant quantities or list the values of certain quantities for various wavelengths in Table \ref{table:Skinproperties}. These values/resources were chosen as they featured data in the wavelength range for our simulations.

\clearpage

\begin{landscape}
\begin{table*}
\begin{longtable}[c]{llll}
\caption{Skin optical properties ($\lambda$ is given in nm).}
\label{table:Skinproperties}\\
\textbf{Tissue} & \textbf{Parameter} & \textbf{Value} & \textbf{Ref.} \\ \hline
\endhead
\hline
\endfoot
\endlastfoot
\multirow{6}{*}{Epidermis} & Optical absorption (cm\textsuperscript{-1}) & $\mu_{ae} = (C_M6.6(\lambda^{-3.33})(10^{11})) + (1-C_M)(0.244 + 85.3(\exp(-\frac{\lambda - 154}{66.2})))$ & \cite{StevenL.Jacques} \\
& Melanosome fraction \textit{C\textsubscript{M}} & 6\% for Caucasian skin, 40\% for pigmented skin & \cite{StevenJacques} \\
 & Reduced scattering (cm\textsuperscript{-1}) & $\mu_s' = 68.7(\frac{\lambda}{500})^{-1.16}$  & \cite{Jacques2013} \\
 & Refractive Index & 1.42 - 1.44 (700 nm - 900 nm) & \cite{Ding2006}  \\
 & Anisotropy & 0.95 - 0.8 (750 nm - 2500 nm)  & \cite{BASHKATOV2011} \\
 & Thickness & 0.1 mm & \\ \hline
\multirow{6}{*}{Dermis} & Optical absorption (cm\textsuperscript{-1}) & $\mu_{ad} = C_B\mu_{ab} + (1-C_B)(0.244 + 85.3(\exp(-\frac{\lambda - 154}{66.2})))$ & \cite{StevenJacques} \\
 & Blood volume fraction \textit{C\textsubscript{B}} & 0.2\% - 7\% & \cite{Yudovsky2011} \\
 & Reduced scattering (cm\textsuperscript{-1}) & $\mu_s' = 45.3(\frac{\lambda}{500})^{-1.292}$ & \cite{Jacques2013} \\
 & Refractive Index & $n = A + \frac{B}{\lambda^2} + \frac{C}{\lambda^4}$, where $A = 1.3696$, $B = 3.9168\textrm{x}10^3$, $C = 2.5588\textrm{x}10^3$ & \cite{Ding2006} \\
 & Anisotropy & 0.95 - 0.8 (750 nm - 2500 nm)  & \cite{BASHKATOV2011} \\
 & sO$_2$ & 40\% - 100\% & \cite{Yudovsky2011} \\ \hline
 \newpage
 Blood & Optical absorption (cm\textsuperscript{-1}) & $\mu_{ab} = C_{Hb}\alpha_{Hb} + C_{HbO_2}\alpha_{HbO_2}$ &\cite{StevenJacques} \\
 & Reduced scattering (cm\textsuperscript{-1}) & $22(\frac{\lambda}{500})^{-0.66}$ & \cite{Jacques2013} \\
 & Refractive Index & 1.36 (680 nm - 930 nm) & \cite{Lazareva2018} \\
 & Anisotropy & 0.994 (Roughly constant for variant wavelength and sO$_2$) & \cite{Bashkatov2005,Faber2004} \\ \hline
Hypodermis & Optical absorption (cm\textsuperscript{-1}) & 1.1 at 770 nm, 1.0 at 830 nm & \cite{Salomatina2006} \\
 & Reduced scattering (cm\textsuperscript{-1}) & 20.7 at 770 nm, 19.6 at 830nm & \cite{Salomatina2006} \\
 & Refractive index & 1.44 (680 nm - 930 nm) & \cite{Bashkatov2005} \\
 & Anisotropy & 0.8 (750 nm - 2500 nm) & \cite{valerytuchin2014} \\ \hline
\end{longtable}
\end{table*}

\end{landscape}


\subsection*{Disclosures}
No conflicts of interest, financial or otherwise, are declared by the authors.

\subsection* {Acknowledgments}
The authors would like to thank Simon Arridge and Paul Beard for helpful discussions. The authors acknowledge support from the BBSRC London Interdisciplinary Doctoral Programme, LIDo, the European Union’s Horizon 2020 research and innovation program H2020 ICT 2016-2017 under Grant agreement No. 732411, which is an initiative of the Photonics Public Private Partnership, the Academy of Finland Project 312123 (Finnish Centre of Excellence in Inverse Modelling and Imaging, 2018--2025) and the CMIC-EPSRC platform grant (EP/M020533/1). 



\bibliography{report}   
\bibliographystyle{spiejour}   






\listoffigures
\listoftables

\end{spacing}
\end{document}